\documentclass[preprint,authoryear,12pt]{elsarticle}

\usepackage{graphicx}
\usepackage{amssymb}
\usepackage{amsmath}

\journal{Astroparticle Physics}

\begin{document}

\begin{frontmatter}


\title{Impact of aerosols and adverse atmospheric conditions on the data quality for spectral analysis of the H.E.S.S. telescopes}

\author[label1]{J. Hahn}
\author[label1]{R. de los Reyes}
\author[label1]{K. Bernl\"ohr}
\author[label1,label2]{P. Kr\"uger}
\author[label5]{Y.T.E. Lo}
\author[label5]{P.M. Chadwick}
\author[label5,label6]{M.K. Daniel}
\author[label1]{C. Deil}
\author[label1,label3]{H. Gast}
\author[label4]{K. Kosack}
\author[label1]{V. Marandon}

\address[label1]{Max-Planck-Institut f\"ur Kernphysik, P.O. Box 103980, D 69029}
\address[label2]{North-West University, Potchefstroom, South-Africa}
\address[label3]{RWTH Aachen University, Germany}
\address[label4]{CEA Saclay, France}
\address[label5]{Durham University, UK}
\address[label6]{Now at University of Liverpool, UK}

\begin{abstract}
The Earth's atmosphere is an integral part of the detector in ground-based imaging atmospheric Cherenkov telescope (IACT) experiments and has to be taken into account in the calibration. Atmospheric and hardware-related deviations from simulated conditions can result in the mis-reconstruction of primary particle energies and therefore of source spectra. During the eight years of observations with the High Energy Stereoscopic System (H.E.S.S.) in Namibia, the overall yield in Cherenkov photons has varied strongly with time due to gradual hardware aging, together with adjustments of the hardware components, and natural, as well as anthropogenic, variations of the atmospheric transparency. Here we present robust data selection criteria that minimize these effects over the full data set of the H.E.S.S. experiment and introduce the \textit{Cherenkov transparency coefficient} as a new atmospheric monitoring quantity.
The influence of atmospheric transparency, as quantified by this coefficient, on energy reconstruction and spectral parameters is examined and its correlation with the aerosol optical depth (AOD) of independent MISR satellite measurements and local measurements of atmospheric clarity is investigated. 

\end{abstract}


\end{frontmatter}


\section{Introduction}\label{}
During the last decade, imaging atmospheric Cherenkov telescopes (IACTs) have qualified as powerful instruments for astronomy in the very-high-energy (VHE; E$>$0.1 TeV) regime, and allowed detailed studies of the most violent phenomena known in the Universe.

The flux of VHE gamma rays is very low, so that large effective areas are required. To achieve this, the IACT technique makes use of telescopes on the ground and the Earth's atmosphere acts as the calorimeter of the detector system.

Due to the atmosphere's opacity, VHE photons can be observed only indirectly at ground level: gamma rays penetrating the atmosphere interact with air molecules and give rise to showers of secondary particles (Extensive Air Showers, EAS). Cherenkov telescopes are designed to detect the  Cherenkov radiation emitted by these relativistic shower particles.
The main strengths of this type of detector, compared to any other ground-based systems, include the good rejection of the background of cosmic-ray initiated showers, the angular and energy resolution, and the low energy threshold.

The energy threshold is directly influenced by the atmospheric absorption, as more absorption leads to a higher threshold.
\\

H.E.S.S. is an array of Cherenkov telescopes situated at Khomas Highland of Namibia (23$^\circ$16'18'' S, 16$^\circ$30'00'' E) at 1800\,m above sea level. Four telescopes are equipped with a 13\,m Davis-Cotton mirror arrangement featuring a focal length of 15\,m. They are located at the corners of a square of 120\,m side length, optimized for an energy threshold of 100\,GeV. Each camera is equipped with an array of 960 photo-multiplier tubes (PMTs) with attached light concentrators, covering a field of view of $5^\circ$ diameter in the focal plane~\citep{Aharonian2006}.

In order to reconstruct the energy of the primary particle, shower images are compared to Monte Carlo shower simulations for which nominal hardware parameters and average atmospheric conditions at the H.E.S.S. site are assumed~\citep{Bernlohr2000}.
However, changes in telescope photon detection efficiency as well as atmospheric variations complicate this comparison. In the case of the telescope photon detection efficiency this includes changes in the photo-sensor response as well as in the reflectivities and transmissivities of optical components such as the telescope mirrors or the light concentrators.

The hardware changes are taken into account through a detailed calibration of the instrument:
\begin{itemize}
  \item PMT aging is monitored by a regular measurement of their gain. Adjustments to the high voltage are performed to keep the gain in a region that is compatible with the Monte Carlo simulations~\citep{Aharonian2004}. 
  \item Changes in telescope photon detection efficiency are measured through the detection of muons during normal data taking. These events allow the determination of the ``muon efficiency'' $\mu$, which is a telescope-wise quantity and provides a measure of the number of photo-electrons detected per incident photon. Although muons are created at any time during the shower development, only those emitted in the last few hundred metres form a ring-shaped image in the camera. By comparing radii and intensities of these rings to theoretical values, it is possible to measure changes in the optical performance~\citep{HESSMuons}. New Monte Carlo simulations are performed when $\mu$ changes by more than $\sim$10-15\%, typically due to changes in mirror reflectivity.
\end{itemize}
While it is possible to measure the effect of these hardware parameters on the spectral shower reconstruction, the effect of the atmosphere is more difficult to quantify, due to its complexity and our limited knowledge of the atmospheric conditions.

Some atmospheric phenomenon will act as atmospheric light absorbers, attenuating Cherenkov light from EAS particles and therefore reducing the amount of Cherenkov photons that reach the detector. Therefore, it is expected that reduction of the actual atmospheric Cherenkov light transparency compared to the Monte Carlo model assumptions result in underestimated energies.

That effect is especially problematic for spectral analysis since misreconstructed energies shift the entire reconstructed spectrum to lower energies. This results in biased values of the reconstructed flux normalization, and, in particular in the case of non-power law spectra, other spectral parameters.

To limit such effects to a minimum, corresponding monitoring quantities have to be used in the Cherenkov technique in order to detect data that is taken  in the presence of clouds and aerosols. 
In this paper we present a new way to estimate the atmospheric transparency by using only observables and calibration parameters from the Cherenkov data taken with the H.E.S.S. telescope array.

In the first part of this paper we will discuss the most important atmospheric conditions that affect spectral shower reconstruction. A second part will present the new monitoring quantity that estimates the ``atmosphere transparency'', followed by a short systematic study on the effect of atmospheric transparency, traced by this quantity, on reconstructed spectral parameters.
Finally, the last part will contain a detailed comparison of this new monitoring quantity with satellite data that measures the total atmospheric optical depth at different wavelengths and with local radiometer measurements of sky clarity.

\section{Origin of atmospheric effects}

\subsection{Clouds}\label{sec:clouds}
The maximum of the Cherenkov emission from air showers, developed by primary particles of energies within the H.E.S.S. energy domain (E$\ge$ 300 GeV), takes place at altitudes between $\sim$6-11 km (see~\citealp{Bernlohr2000}). Thus, one can assume that thin layers of clouds below those altitudes act as atmospheric light absorbers which may attenuate Cherenkov light from the whole shower or parts of it, resulting in fewer photons reaching the camera and a lower trigger probability. As a result, single telescope trigger rates and consequently the central trigger rate (see~\citealp{Funk2004}) are reduced\footnote{The trigger rate decreases with the zenith angle even in absence of atmospheric light absorbers due to the increase of the distance of the shower maximum. In this paper, trigger rates are assumed to be corrected for this effect.}. The trigger rate can therefore be used to detect data that has been taken in the presence of clouds.

\begin{figure}[h!!!]
  \begin{center}
  \includegraphics[width=\textwidth]{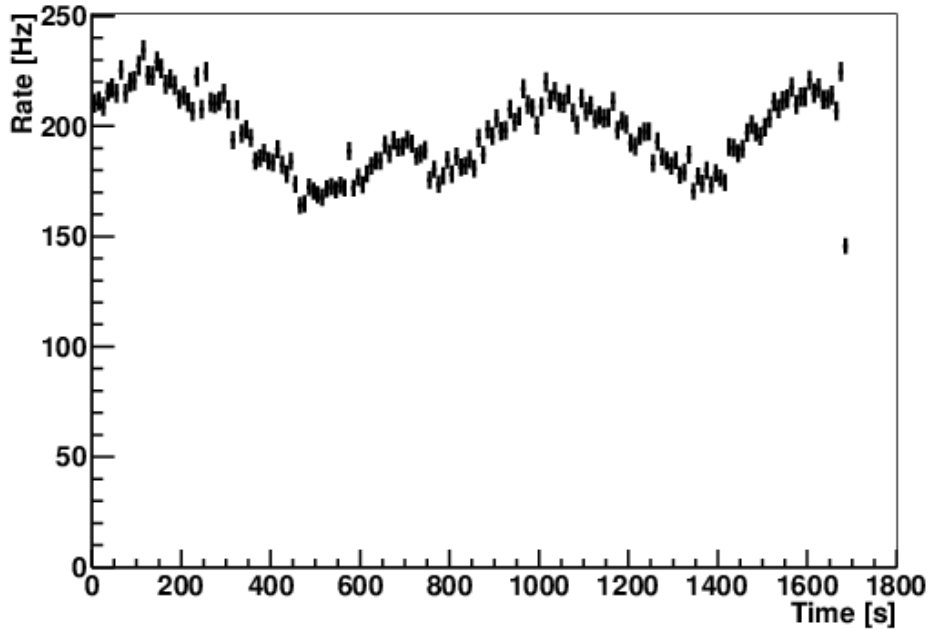}
  \caption{Fluctuating behavior of the central trigger rate in the presence of clouds moving through the field of view. Fluctuations can be quantified by the RMS of the data points with respect to a linear fit.}
  \label{figure0}
  \end{center}
\end{figure}

For instance, if absorbing structures (local clouds) pass through the field of view, a fluctuating behavior in the central trigger rate on time scales smaller than the standard duration of data sets\footnote{In H.E.S.S., observations are performed in so-called runs of 28 minutes.} can be observed, as highlighted in Fig.~\ref{figure0}. 
In order to quantify these fluctuations for each run, the average central trigger rate (coincidence of two or more telescopes) is calculated over 10s time intervals. The resulting evolution of the trigger rate is fitted by a linear function. Fluctuations in the central trigger rate can be quantified by the R.M.S. of the residuals of the fit. As an atmospheric data monitoring quantity, the R.M.S. value is divided by the time averaged central trigger rate. In the absence of clouds, relative fluctuations in the central trigger rate are smaller then 3\%.

However, this quantity is only sensitive to clouds that affect the central trigger rate on time-scales smaller than the run duration. Long-term atmospheric absorbers like aerosol and long-term cloud layers need another quantity to be detected. 

\subsection{Aerosols}
In addition to clouds, there is another phenomenon that might absorb or scatter photons from Cherenkov showers: aerosol particles of human or natural origin.
Even though the atmosphere at the H.E.S.S. site features a very low aerosol concentration, seasonal biomass burning to the north-east of the H.E.S.S. site around August and October may lead to a significant increase of the atmospheric aerosol content.
These aerosols can be transported over large distances to the H.E.S.S. site resulting in an increase of the aerosol concentration over several kilometres in height at the site (see~\citealp{Bernlohr2000}).
The aerosols may also persist in the atmosphere for weeks, affecting the experiment over time scales much larger than that of individual data sets. 
This makes it difficult to disentangle their effect on the trigger rates from that of instrumental changes, including those resulting from maintenance work and adjustments of PMT voltages.

\section{Cherenkov transparency coefficient}
In order to detect data that have been taken in the presence of elevated aerosol concentrations and of large-scale clouds, we have developed a new quantity, the \textit{Cherenkov transparency coefficient}. This quantity is designed to be as hardware-independent as possible in order to separate hardware-related effects from the decrease in trigger rates caused by large-scale atmospheric absorption.

The transparency coefficient is derived under the assumption that the zenith-corrected single telescope trigger rates $R$ are dominated by cosmic-ray (CR) protons.
The local CR proton spectrum in the relevant energy range is approximately $f(E) = 0.096\cdot (E/\text{TeV})^{-2.70} \text{m}^{-2} \text{s}^{-1} \text{TeV}^{-1}\text{sr}^{-1}$ ~\citep{BESS98}.
Hence, the trigger rates can be estimated by
\begin{eqnarray}
	R &\sim& \int_{0}^\infty \text{d}E A_{\rm eff}(E)f(E)\\ 
	  &\simeq& k\cdot E_0^{-1.7 + \Delta},
\end{eqnarray}
where $E_0$ is the energy threshold of the telescopes. The term $\Delta$ allows to take into account higher order corrections, such as energy-dependent shower profiles.

Furthermore, $E_0$ is assumed to be inversely proportional to the product of the average pixel gain $g$ (see~\citealp{Aharonian2004}) and the telescope-wise muon efficiency $\mu$~\citep{HESSMuons}.  The atmospheric transparency is parametrized by a factor $\eta$ and assumed also to be inversely proportional to $E_0$ so that $ E_0 \propto (\eta \cdot \mu \cdot g )^{-1} $.
Since $\mu$, $g$ and $R$ are observables or calibration parameters it is possible to calculate $\eta$ for each telescope using 
$$\eta \propto \frac{R_i^{\frac{1}{1.7-\Delta}}}{\mu_i\cdot g_i} \equiv t_i.$$

The product of muon efficiency with average pixel gain corrects for gain re-adjustments, rendering the muon efficiency sensitive only to degradations of the mirrors and funnels.
It should be noted also that instead of the single telescope trigger rates the telescope read-out rates are used. Telescopes are read out when at least two telescopes are triggered in coincidence (see~\citealp{Aharonian2004}). The coincidence requirement suppresses the effects of random fluctuations present in the single telescope trigger rates and improves the stability of the quantity. However, it also introduces a correlation between the telescope rates. Furthermore, due to the coincidence requirement, the single telescope read-out rates depend on the number of active telescopes.

\begin{figure}[h!!!]
  \includegraphics[width=1.\textwidth]{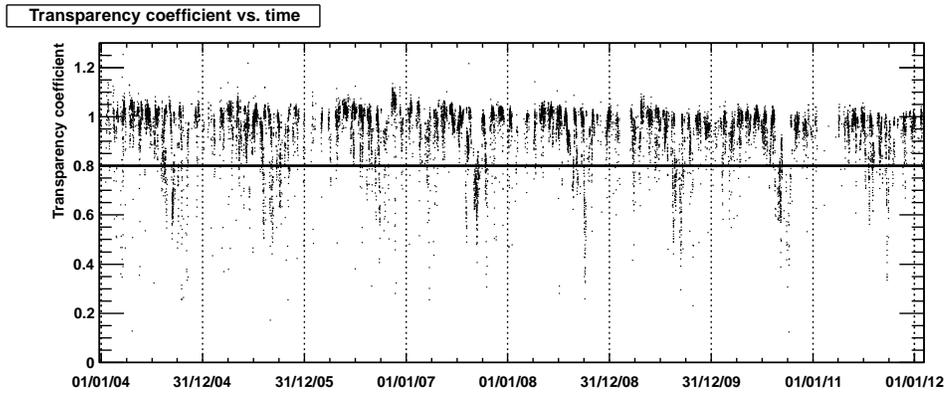}
  \caption{Evolution of transparency coefficients over the last 8 years.  The solid line indicates the cut value at 0.8. The distribution is sharply peaked at 1 with a FWHM $\sim$ 9$\%$.}
  \label{figure1}
\end{figure}

To get a quantity for the whole system that is independent of the number of active telescopes, the average of $t_i$ over all $N$ active telescopes is calculated and rescaled by a factor of $k_N$ so that the corresponding distribution peaks at a value of one.
The factor $k_N$ depends on the telescope multiplicity due to the coincidence requirement of the telescope read-out. It is determined by the peak positions of the $\langle t \rangle_N = \sum_i{t_i/N}$ distribution of the whole data set between the years 2008 and 2012. The numerical values are $k_3 = 3.11$ and $k_4 = 3.41$. The normalization to the peak value will cancel the contribution of other CR species.
The {\it Cherenkov transparency coefficient} is then defined as
$$T \equiv \frac{1}{N\cdot k_N}\sum_i{t_i}.$$
The T values, measured over an 8 year period, are displayed in Fig.~\ref{figure1} assuming $\Delta = 0$. For the whole period, the peak values are found to be close to 1, illustrating that the atmospheric transparency measure T is indeed independent of hardware effects.


\begin{figure}[h!!!]
  \includegraphics[width=0.5\textwidth]{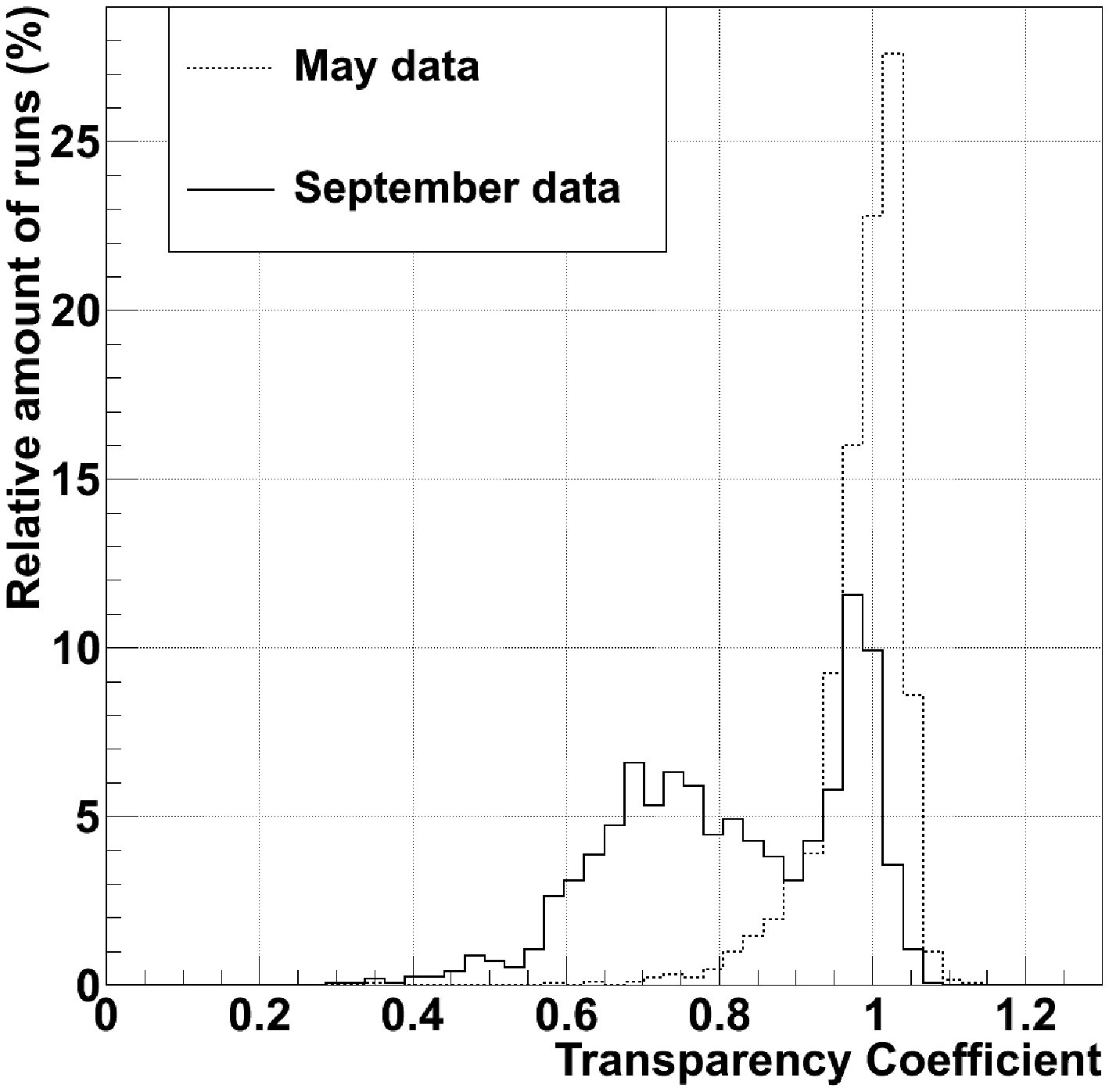}
  \includegraphics[width=0.5\textwidth]{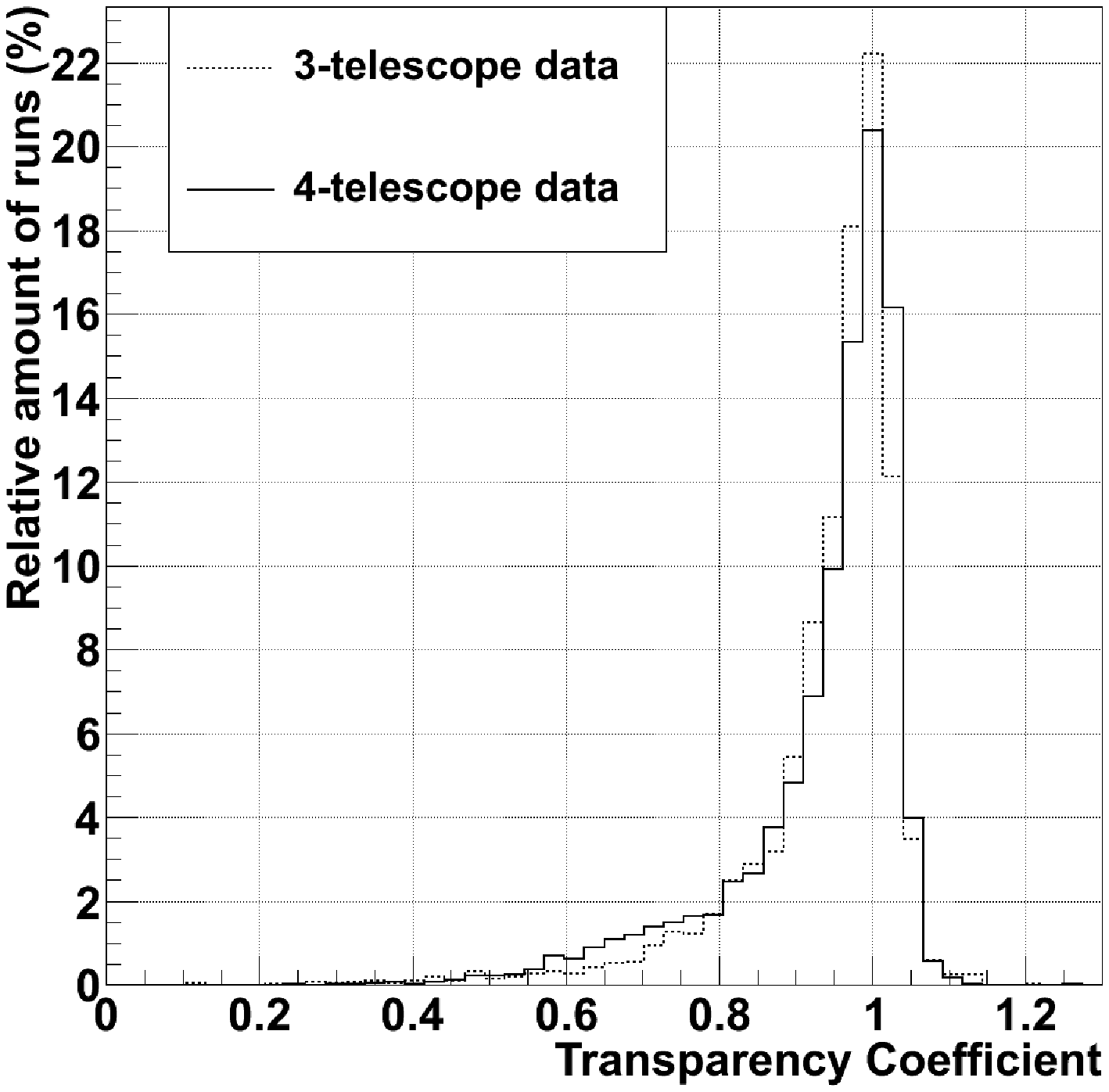}
  \caption{Left: Distribution of transparency coefficients for data taken during the months of May and September between 2004 and 2011. Right: Distribution of transparency coefficients for all-year data during the same 8 year time period taken with 3 and 4 telescopes, respectively.}
  \label{figure2}
\end{figure}

In the right panel of Fig.~\ref{figure2} the distribution of transparency coefficients is shown for both 3- and 4-telescope data for the month of May. As one can see, the distributions are narrowly peaked around a value of one.
 
The left panel of Fig.~\ref{figure2} shows the complete (3- and 4-telescope data) distribution of transparency coefficients for both May and September data. The peak close to a value of one is present in both distributions, indicating the abundance of good atmospheric conditions in May as well as September. However, a second, much broader peak in the distribution of transparency coefficients during September can be observed. This additional feature corresponds to the periodic downward fluctuations that can be seen in Fig.~\ref{figure1} and is coincident with periods of biomass burning in Namibia and its neighbouring countries.

\section{Systematic effects on reconstructed spectra}
The correlation between the transparency coefficient and spectral distortions was studied on the example of the Crab Nebula, a standard candle at TeV energies without any detectable variability over timescales of years. The data were taken over 8 consecutive years from 2004 to 2011. The standard quality criteria, with the exception of those for good atmospheric conditions, were applied to remove runs with technical problems or runs affected by small clouds in the field of  view. The full data set was subdivided into subsets of data corresponding to different ranges of the atmospheric transparency parameter. The standard cut-based analysis using simple air shower image parameters, the so called Hillas analysis ~\citep{Hillas1985}, was then employed to obtain spectral information for each subset.

The gamma ray spectrum of the Crab Nebula has been measured by H.E.S.S. ~\citep{Aharonian2006} and was found to have an approximate power-law shape with some bending at the highest energies. For a pure power-law fit in the energy range ($0.41$-$40$)TeV, the flux normalization at 1 TeV, $\phi_{0,Crab}$, and the spectral index, $\Gamma_{Crab}$, were found to be $\phi_{0,Crab}=(3.45\pm0.05_{stat}\pm0.69_{sys})\times 10^{-11}\text{cm}^{-2}\text{s}^{-1}\text{TeV}^{-1}$ and $\Gamma_{Crab}=2.63\pm0.01_{stat}\pm0.10_{sys}$.

\begin{figure}
  \begin{center}
      \includegraphics[width=.49\textwidth]{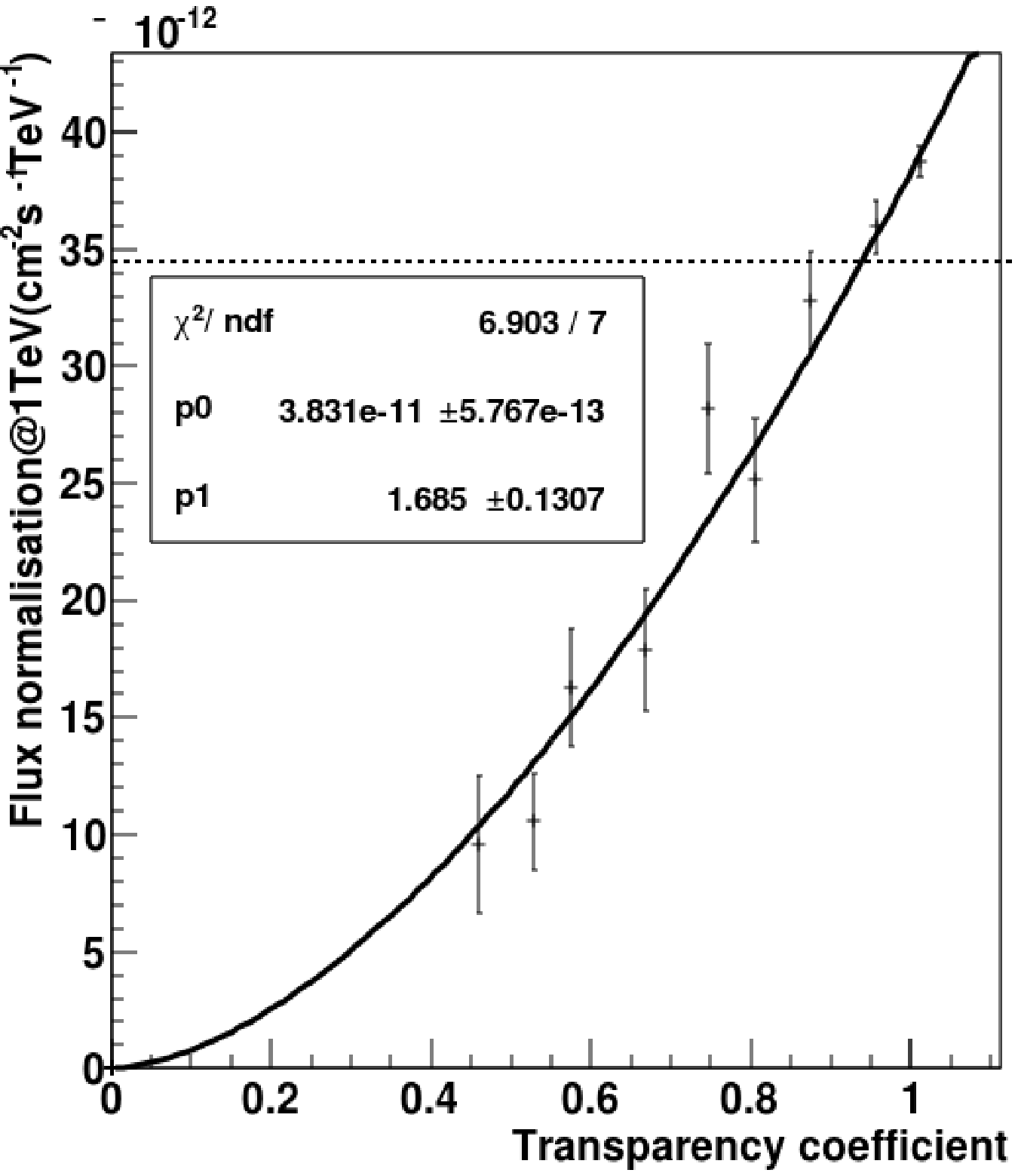}
      \includegraphics[width=.49\textwidth]{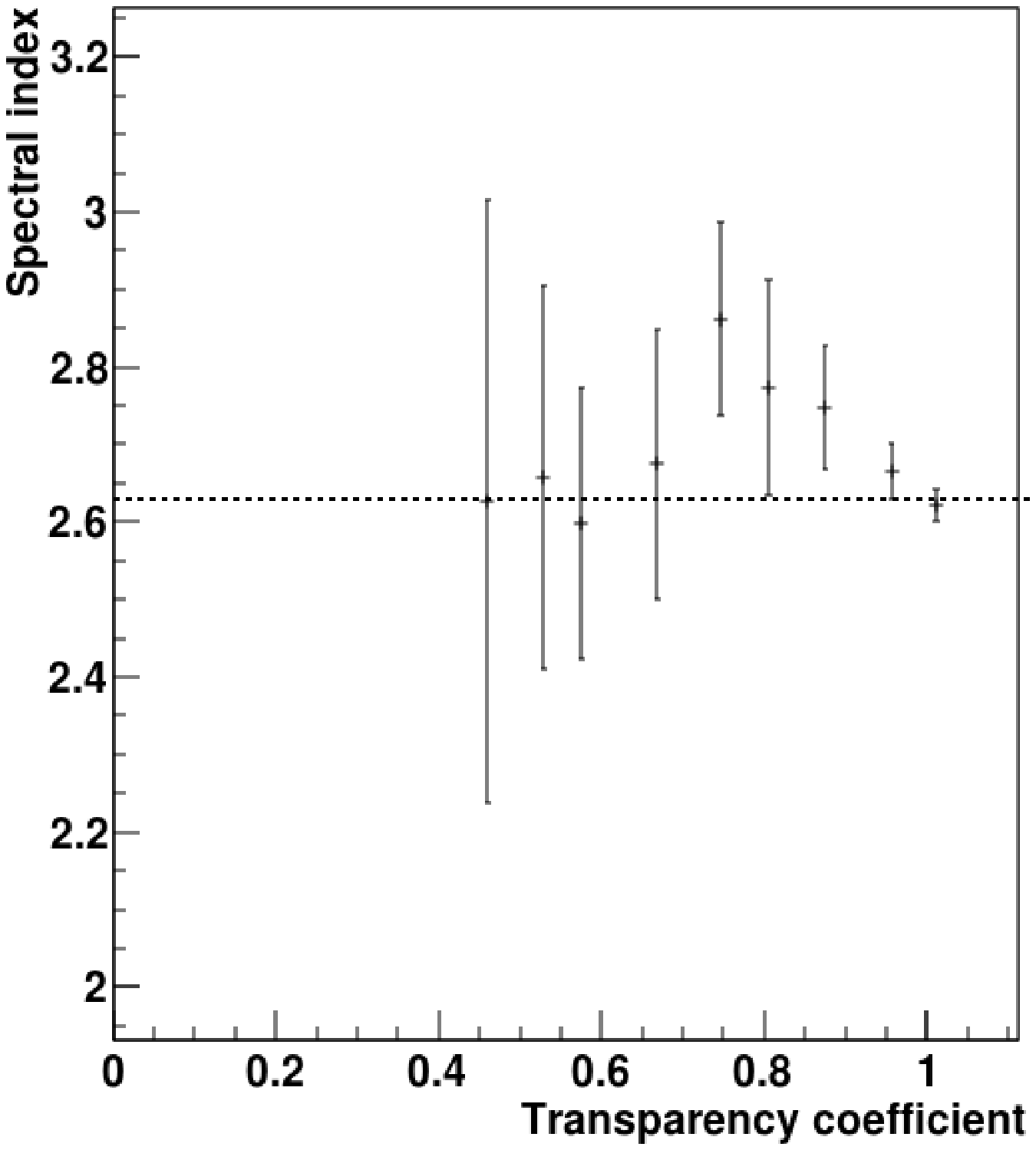}
  \end{center}
  \caption{Flux normalization at 1 TeV (left) and spectral index (right) for Crab Nebula data taken during 8 years of H.E.S.S. operation. Abscissa values are given by the mean value of the transparency coefficient in the respective subset. In the left panel, best fit values for a power-law fit are shown for the flux normalization at T=1 (p0) and the exponent (p1). The full data set investigated has an exposure of 84 hours, using only observations within one degree offset from the source. Also, to minimize a possible zenith-dependent energy bias, only data taken at zenith angles smaller than 47 degrees have been selected. The analysis performed uses the Hillas method where standard cuts~\citep{Aharonian2006} have been applied. Dashed lines represent the published results (see~\citealp{Aharonian2006}).}
  \label{figure3}
\end{figure}

Assuming that atmospheric absorption leads to an underestimation of the reconstructed energy by a constant attenuation factor, a power law spectrum is expected to stay form invariant with changing atmospheric conditions. However, this shift in the spectrum energy range is expected to bias the estimated flux normalization at a given reconstructed energy. Quantitatively, assuming the reconstructed gamma ray energy $E_{reco}$ and the true energy $E_{true}$ to be related via $E_{reco}\propto T\times E_{true}$, 
one finds
\begin{eqnarray} \frac{\text{d}F}{\text{d}E_{true}} \propto E_{true}^{-\Gamma} ~~~~~\Leftrightarrow ~~~~~ \frac{\text{d}F}{\text{d}E_{reco}} \propto E_{reco}^{-\Gamma}\cdot T^{\Gamma-1}
\label{eqfit}
\end{eqnarray}
Fig. \ref{figure3} (left panel) displays the reconstructed flux normalization $\phi_0$
as function of the measured transparency coefficient $T$, confirming the
expected strong T-dependence. Fitting a power law function $\propto T^{\Gamma-1}$
as expected from (\ref{eqfit}) yields a good description of the data
(solid line in Fig. \ref{figure3}, left panel) with $\Gamma_{fit}=2.69\pm 0.13$, in perfect
agreement with the spectral index obtained from the power law fit to the
Crab Nebula spectrum ($\Gamma_{Crab}$). 

Fig. \ref{figure3} (right panel) displays the reconstructed spectral index $\Gamma$
as function of $T$. An indication for some hardening of the spectrum with increasing T in the range $0.7\le T\le 1$ might be observed, but it is not significant. 
This may be at least partially attributed to deviations of the spectrum of the Crab Nebula from
a power law which leads to an increase of the spectral index for
increasing energy threshold (connected to decreasing transparency).
Also variations of the
shower shapes with energy may contribute to this behaviour but, since the T-dependence is found to be weak, we conclude that the simple attenuation model 
(\ref{eqfit}) is at least a good approximation.

The sensitivity of the transparency coefficient to reduced reconstructed flux values, connected to a decreased atmospheric transparency, allows it to be used as a monitoring quantity in data quality selection. In the presented example and for the data range above a transparency coefficient value of 0.8, relative flux variations (to the mean flux in that range) are limited to about 20\%.
 
\section{Aerosol atmospheric absorption}

The relation of the Cherenkov transparency coefficient with the concentration of aerosols should be proven through a positive correlation with independent aerosol measurements such as satellite measurements. In the following we will search for a correlation between the Cherenkov transparency coefficient and the Aerosol Optical Depth (AOD)\footnote{The aerosol optical depth or optical thickness is defined as the integrated aerosol extinction coefficient over a vertical column of unit cross section.}; or more specifically with the atmospheric transparency ($\propto \exp(-\mathrm{AOD})$).

The influence of the atmospheric aerosols on transparency is quite complex and strongly depends on the detailed scattering and absorption properties of the different aerosol types (e.g. sulphate, dust, organic carbon, sea salt), which might yield differences in atmospheric transparency of up to 10\% in the UV-B band~\citep{Balis2004}.

Many studies of the atmospheric absorption of aerosols have been carried out, not only for astronomical purposes but also for climate and atmospheric studies.
However, we are particularly interested in those related with increases of aerosol absorption due to biomass burning.

\subsection{Previous results of satellite data}

Previous studies of MODIS (Moderate Resolution Imaging Spectroradiometer) satellite data have been made at other observatories~\citep{Arola2007} to measure surface radiation levels during periods of biomass burning in the surroundings.
These biomass burning aerosols seem to decrease the amount of UV solar radiation reaching the surface by up to 50\%, with typical values in the range of $\sim$ 15-35\%~\citep{Kalashnikova2007}. 

The MISR (Multi-angle Imaging SpectroRadiometer) instrument, on board NASA's Terra spacecraft, has a better grid spatial resolution (1.1 km in global mode)~\citep{Diner1988} which was used to confirm the previous MODIS results. With the capability of observing at different viewing angles, MISR can distinguish between different types of atmospheric particles (aerosols), different types of clouds and different land surfaces. The processed (Level 3) AOD data have proven to be in better agreement with the ground-based Aerosol Robotic Network (AERONET) measurements~\citep{Tesfaye2011} than previous satellite measurements.

In particular, a detailed 10-year study of the aerosol climatology with MISR over South Africa, Namibia's neighbour country, has revealed that the northern part of South Africa seems to be rich in aerosol reservoirs and the aerosol concentration (based on optical depth) is 34\% higher than that in the southern part of the country~\citep{Tesfaye2011}. This study proves the occurance of biomass activity not so far from the H.E.S.S. site, hence an effect on the Cherenkov light (mainly in UV wavelengths) from the EAS particles is expected.

\subsection{Correlation between Cherenkov transparency coefficient and MISR data}

~\citealp{Tesfaye2011} have also found seasonal changes in the aerosol composition in South Africa. During summer and early winter (in the southern hemisphere), the northern part of South Africa is dominated by a mixture of coarse-mode and accumulation-mode particles (i.e. particles of around 1-20$\mu$m radius), which are a result of air mass transport from arid/semi-arid regions of the central parts of South Africa, Botswana and Namibia. In the time from August to October (winter and early summer) it is dominated by sub-micron particles. The most important sources of sub-micron particles are industrial and rural activities (including mines and biomass burning).

The periodic drops in the Cherenkov transparency coefficient for the H.E.S.S. site (see Fig.~\ref{figure1}) correlate with the seasonal increase of sub-micron particles due to, among others, biomass burning like in the neighbouring South Africa. This gives an indication of the main atmospheric phenomenon responsible for the reduced trigger rates of some H.E.S.S. observations, especially in early summer.

As we have mentioned before, if the Cherenkov transparency coefficient is a good data quality parameter to monitor the atmospheric transparency, we expect a strong and positive correlation with the AOD measured by satellites.
To carry out such a comparison, the AOD retrieved from MISR data and the Cherenkov transparency coefficient from H.E.S.S. data were used. Both data sets cover the same period of time between 2004 and 2011. 

The processed (level 3) MISR AOD data at the H.E.S.S. site (with a grid spacial resolution of 0.5$^\circ$x0.5$^\circ$) at three different wavelengths (443 nm, 555 nm and 670 nm), from UV to red wavelengths were used. Note that the satellite only measures the AOD during daytime.
On the other hand, the values of the Cherenkov transparency coefficient are calculated in intervals of 28 minutes during the dark time period at the H.E.S.S. site. The Cherenkov transparency coefficient data set has been filtered by data quality cuts that remove data taken during cloudy nights, as already described in Section~\ref{sec:clouds}, and hardware problems. 
 
Depending on latitude, the satellite samples a fixed location every 2 to 9 days. The overlap of satellite measurements and H.E.S.S. data taking is therefore sparse, reducing the available data set for the correlation study (in this study, only MISR data that has been taken within 24 hours of H.E.S.S. measurements is used, corresponding to 2\% of the whole H.E.S.S. data set).

\begin{figure}
  \includegraphics[width=1.\textwidth]{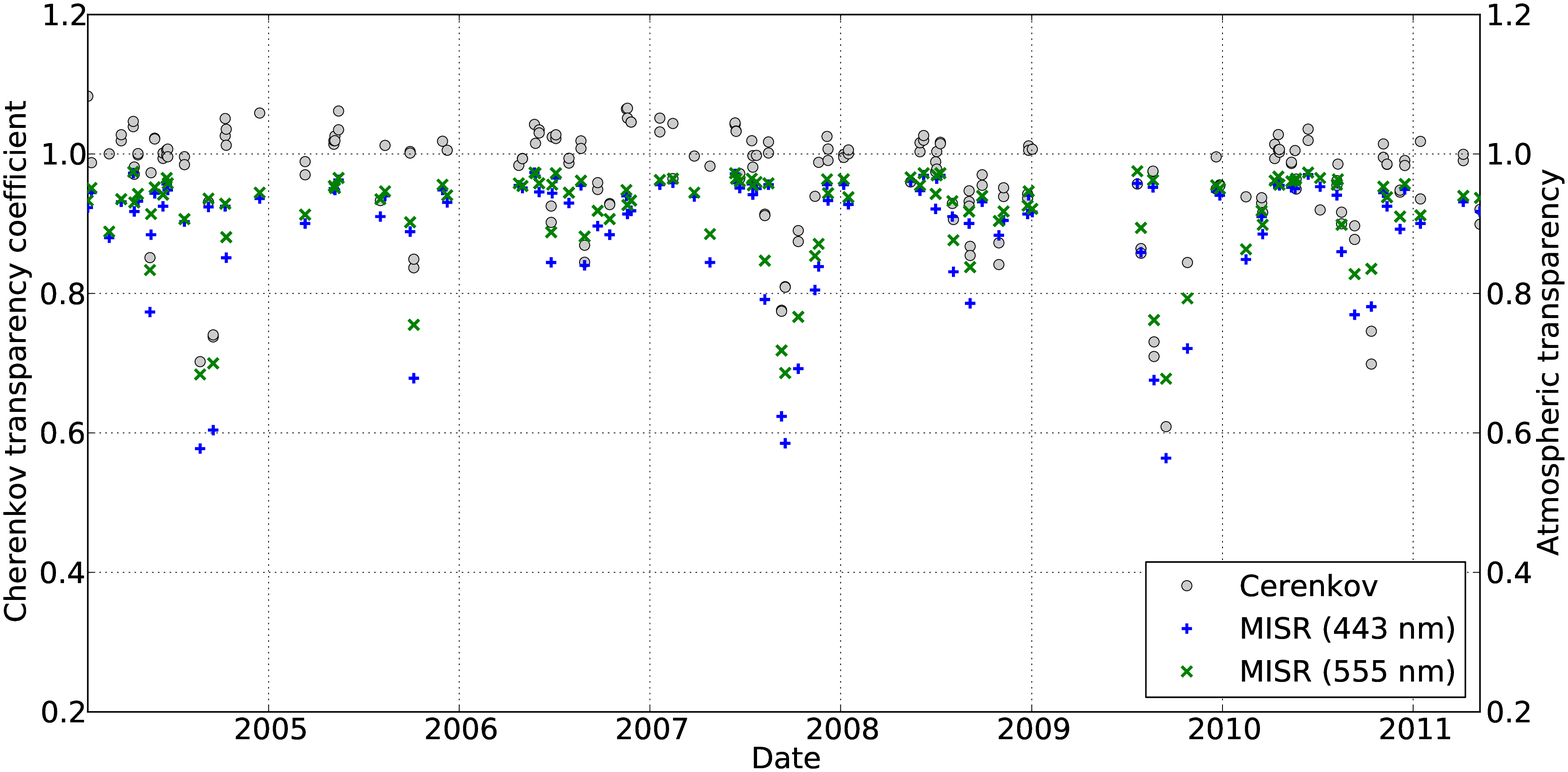}
  \caption{Cherenkov transparency coefficient measured in the time interval between 2004 and 2011, together with the MISR atmospheric transparency measurements in 443 nm (blue points) and 555 nm (green points).}
  \label{figure5}
\end{figure}

The correlation of the Cherenkov transparency coefficient with the MISR-retrieved atmospheric absorption at blue and green wavelengths can be seen in Fig.~\ref{figure5}, which shows the temporal evolution of MISR-AOD and the Cherenkov coefficient. 

\begin{figure}
  \includegraphics[width=1.1\textwidth]{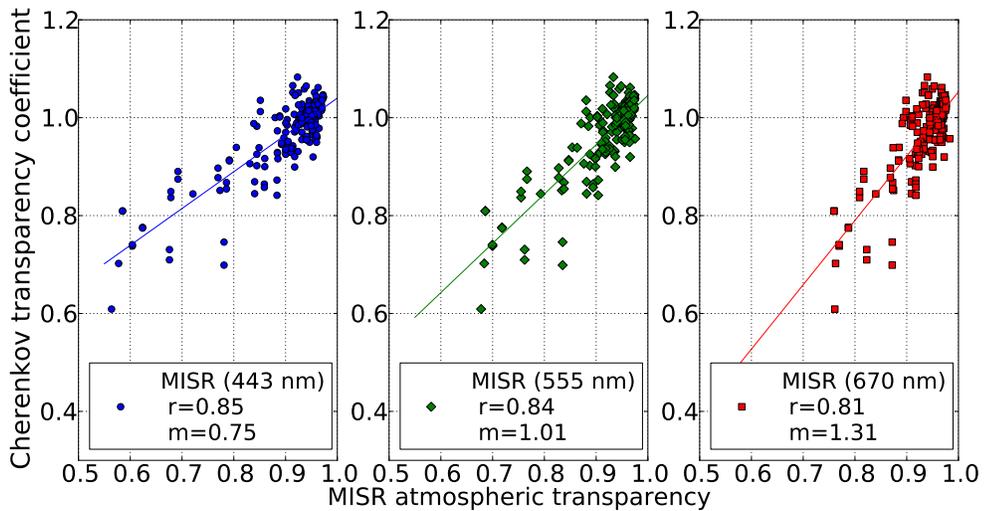}
  \caption{MISR atmospheric transparency ($\exp(-\mathrm{AOD})$) against the Cherenkov transparency coefficient. The three wavelengths measured by the MISR satellite are represented in different colors: 443 nm, blue; 555 nm, green and 670 nm, red. The resulting correlation is plotted as a solid line with the corresponding color of the MISR wavelength.}
  \label{figure4}
\end{figure}

Figure~\ref{figure4} quantifies the correlation between the atmospheric absorption ($\propto \exp(-\mathrm{AOD})$) for the three different wavelengths measured by MISR (443 nm in blue, 555 nm in green and 670 nm in red) and the Cherenkov transparency coefficient. The solid lines are the results of a linear fit between measurements ($T = m \cdot \exp(-\mathrm{AOD}) + c$).
The Pearson's correlation coefficients for the wavelengths 443 and 555 nm (blue and green) are $\sim$0.85 and $\sim$0.84 respectively. This shows a positive and strong correlation between the 
atmospheric transparency measured from satellites and the Cherenkov transparency coefficient. The correlation factors for the different wavelengths are identical within errors, although one might expect a stronger correlation in the blue band, since the number of Cherenkov photons emitted per path length in a certain wavelength range (eq. (1) in ~\citealp{Bernlohr2000}) is maximal in the UV-blue part of the spectrum.

Figure~\ref{figure4} also shows an increase of the steepness (``m'' in the figure) of the best fit of the linear correlation, with increasing wavelength. This is due to the fact that the atmospheric transparency measured with the MISR satellite decreases towards shorter wavelengths, while the Cherenkov transparency coefficient is always the same.
For blue wavelengths (443nm; blue line), the linear best fit yields a slope of m=0.75 increasing to m=1.01 at larger wavelengths (555 nm; green line). If the Cherenkov transparency coefficient and the blue MISR transparency would measure the identical atmospheric absorption, one would expect a slope of m=1 for blue, m=$\alpha_b/\alpha_g$ for green and m=$\alpha_b/\alpha_r$ for red, where $\alpha_x$ are the absorption coefficients at the respective colors and $\alpha_b > \alpha_g > \alpha_r$. An increase of the optical depth in the UV-B band, relative to larger wavelengths, was also found by~\citep{Kalashnikova2007} in a study of aerosols resulting from fires in Australia.

The decrease of the atmosphere transmission with decreasing wavelength can be explained by simple Mie scattering.~\citet{Tesfaye2011} established an inverse proportionality between the aerosol particle size and their extinction efficiency at a certain wavelength. An increase of the AOD at short wavelengths therefore indicates the presence of sub-micron (radii$<$0.35 $\mu$m) particles, attributed by the authors to urban pollution (sulphates) and extensive biomass burning activities (carbonaceous aerosols). As a consequence, the aerosol-induced reduction in the atmosphere transparency is expected to be more pronounced at shorter wavelengths, which is where the bulk of the Cherenkov light is emitted.

\subsection{Correlation between Cherenkov transparency coefficient and radiometer temperature}

At the H.E.S.S. site, a number of instruments are used to monitor atmospheric clarity. One of these is a LIDAR (Light Detection And Ranging), centred at 355 and 532~nm, which has been operated in conjunction with the telescopes since mid-2011. The LIDAR backscatter signal can be used to determine atmospheric transmission, and is particularly useful for the detection of aerosols. However, the LIDAR measurements have some limitations. First, this LIDAR is effective only above $\sim$~1~km due to geometric effects. Second, as the LIDAR must operate at a wavelength close to that of the Cherenkov light in order to provide a transmission profile that is relevant to the telescope observations, it can contaminate those observations. Thus, the LIDAR can be operated only between observations and the data are not strictly contemporaneous with the telescope data.

In addition to the LIDAR, a number of Heitronics KT19.82 radiometers are operated at the H.E.S.S. site, with each telescope having a paraxially mounted radiometer and one being used to scan the sky continuously. These radiometers operate between 8 and 14~$\mu$m, in which region clouds are efficient radiators. Radiometers have been shown to be excellent cloud monitors, and the sky temperature correlates well with Cherenkov telescope count rates~\citep{Buckley}. Aerosols are also expected to contribute to the bolometric luminosity of the sky at the level of 30~Wm$^{-2}$~\citep{Dalrymple}, and preliminary studies indicated that the H.E.S.S. radiometers are indeed sensitive to the presence of aerosols~\citep{Daniel}.

\begin{figure}
  \begin{center}
      \includegraphics[width=.8\textwidth]{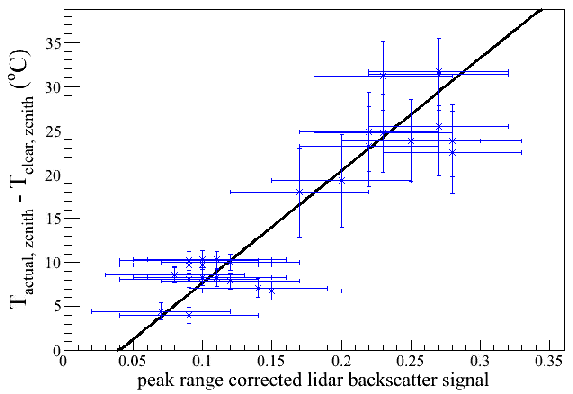}
      \includegraphics[width=.8\textwidth]{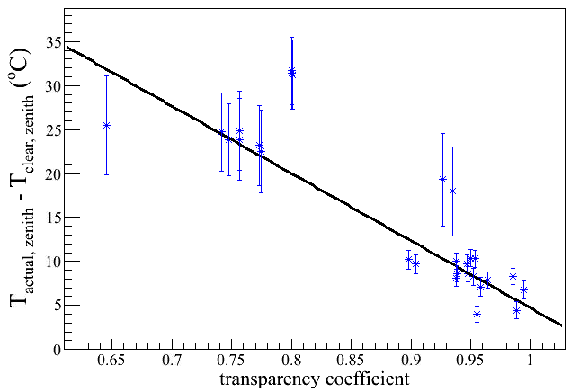}
  \end{center}
  \caption{Results of the analysis of 28 observations which showed evidence for the presence of aerosols and the absence of clouds. Top: The difference between the measured zenith-corrected radiometer temperature and that which would be expected for a clear sky, plotted against the peak range-corrected LIDAR signal, showing that the radiometer temperature is correlated with the density of the aerosol layer. Bottom: The difference between the measured zenith-corrected radiometer temperature and that which would be expected for a clear sky, plotted against the Cherenkov transparency coefficient.}\label{figure7}
\end{figure}

We present here the results of a further study of the sensitivity of radiometers to aerosols. 

Radiometer data taken during the biomass burning period in Namibia between July 2011 and March 2012 were examined for runs which showed, from the most nearly-contemporaneous LIDAR profiles, evidence for the presence of an aerosol layer and the absence of cloud. Some data quality cuts were also applied based on trigger rate, number of telescopes operating, and the quality of LIDAR and radiometer data. This resulted in 28 4-telescope runs for analysis. Since radiometer temperature is correlated with zenith angle and with air temperature, it is usual in studies of radiometer data to correct for both effects. A plot of the peak range-corrected LIDAR backscatter signal, which represents the peak aerosol density, versus the difference between the zenith-corrected sky temperature measured by the radiometer and that which would have been expected for a clear sky, is shown in Figure~\ref{figure7}. The Pearson's correlation coefficient for these data is $\sim~0.92$ indicating a strong correlation between the radiometer temperature and the presence of aerosols and showing that radiometers are a useful tool for measuring aerosols.

Figure~\ref{figure7} further shows the correlation between the Cherenkov transparency coefficient for the 28 aerosol-contaminated observations analysed and the radiometer temperature. Here again, a strong correlation is observed, with the Pearson's correlation coefficient of $\sim~-0.86$. This provides further confirmation of the sensitivity of the Cherenkov transparency coefficient to the presence of aerosols. 

\section{Conclusions}

Changes in the central trigger rates of the H.E.S.S. array are caused both by the changes in telescope properties and in atmospheric conditions.
Observations where layers of clouds pass through the field of view can be identified relatively easily by calculating the relative fluctuation of the central trigger rates.

However, disentangling atmospheric extinction effects that take place on longer time scales (of the order of weeks, e.g. caused by aerosols) from long-term trigger rate decreases caused by a reduced photon detection efficiency connected to the aging of the instrument is an issue (see~\citealp{HEGRA},~\citealp{Chadwick} and~\citealp{VERITAS}).

The situation is further complicated by upgrades and maintenance work, inducing rapid changes in the behaviour of the instrument. To deal with these challenges, a hardware-independent data quality parameter, the \textit{Cherenkov transparency coefficient}, has been developed.

This quantity measures the atmospheric transparency and is sensitive to an increase of absorber concentrations. Also, we have proven that it is stable to within 9\% over the last 8 years of H.E.S.S. data, illustrating its independence from hardware-related effects, like PMT high-voltage adjustments and mirror refurbishments. Furthermore, it is calculated using only basic observables and calibration parameters, which renders it a generic quantity for the Cherenkov technique.

Air showers that have been attenuated by atmospheric absorbers may be systematically mis-reconstructed to lower energies. A systematic study has been performed in order to study the influence of the atmosphere transparency, as measured by the Cherenkov transparency coefficient, on the gamma-ray spectral reconstruction. It was found that the shift in reconstructed energies leads to flux normalizations that may be biased on the order of tens of percent if the data was taken in the presence of thin clouds and/or aerosols. Applying the presented quality cuts to Crab Nebula data largely removes bad-weather data and limits biases in the flux normalization caused by these atmospheric light absorbers to less than 20\%.


A correlation study between the Cherenkov transparency coefficient and AOD measurements performed by the MISR satellite results in a correlation factor of about 0.85 at blue wavelengths ($\lambda$=443 nm). 

A large correlation factor proved the relation (with gradient = 0.75 at 443 nm) between AOD values and the new transparency coefficient, a quantity which is extracted directly from the Cherenkov technique. Our results are in good agreement with previous satellite studies that investigated other regions in the vicinity of the H.E.S.S. site which are known to feature aerosol increases due to human/industrial activities in their surroundings.

In addition, the Cherenkov transparency coefficient is shown to correlate well with local measurements of aerosols provided by radiometers operated at the H.E.S.S. site, with a correlation factor of -0.86.\\
Although there is a good correlation between these measures of aerosols and the Cherenkov transparency coefficient, there are several factors that might limit the correlation: 
\begin{itemize}
  \item The Cherenkov coefficients are based on some simplified assumptions, such as the perfect inverse proportionality between telescope energy threshold and the muon efficiency.
  \item The coefficient does not allow disentangling large-scale thin clouds from layers of aerosols.
  \item No data are taken when the trigger rate is very low due to large AODs. This results in a small amount of data being available at low values of the Cherenkov transparency coefficient.
  \item The time difference between the satellite and H.E.S.S. measurements restricts the temporal accuracy of the correlation to a time-scale of one day.
\end{itemize}
Future work should result in a better understanding of the correlation between atmospheric absorption measured by satellites and the Cherenkov transparency coefficient by addressing the above points. For example, a better disentangling in the Cherenkov data between atmospheric absorption due to aerosols and due to long-term high clouds might improve the correlation. 
Simultaneous observations of on-site LIDAR data and the Cherenkov telescope might also help and are currently under study.

While the Cherenkov transparency coefficient is currently used as a data quality parameter in H.E.S.S., the strong correlation with independent atmospheric measurements suggests that it could also be used to correct the flux for changes in atmospheric conditions. Previous methods, using other atmosphere-sensitive parameters, have already applied this idea (see~\citealp{HEGRA},~\citealp{VERITAS},~\citealp{MAGIC} and~\citealp{Chadwick}).

A similar application of the Cherenkov transparency coefficient is currently under investigation. This would make it possible to use the Cherenkov technique over a wider range of atmospheric conditions.

The calculation of the Cherenkov transparency coefficient for other IACTs could be implemented since all the parameters needed are calculated in their routine calibration and quality checks. Arrays consisting of Cherenkov telescopes with small ($\varnothing$$<$ 5 m) size dishes or of one single dish might have more constrains due to the limited muon statistics in normal observation runs (see ~\citealp{HEGRA}) and the fluctuations in the trigger rate, respectively.
Its implementation in the future CTA is under study, since a large number of large ($\varnothing$$\sim$ 23 m) and medium ($\varnothing$$\sim$ 12 m) size Cherenkov telescopes will be always present in the array.

\section*{Acknowledgements}

The authors would like to acknowledge the support of their host institutions. We want to thank the H.E.S.S. collaboration for their support, especially Prof. Werner Hofmann and Prof. Christian Stegmann as well as Prof. Thomas Lohse and Dr. Ira Jung for the many fruitful discussions.


\bibliographystyle{elsarticle-harv}

\end{document}